\documentclass[12pt]{article}
\usepackage{graphicx}
\usepackage{natbib}
\usepackage{amssymb}

\begin{document}

\title
{Numeric Deduction in Symbolic Computation. Application to
Normalizing Transformations}
\author
{Ivan~I.~Shevchenko
\\ Institute of Theoretical Astronomy,
Russian Academy of Sciences, \\
Nab.~Kutuzova 10, St.Petersburg 191187, Russia}

\date{}

\maketitle

\begin{abstract}
Algorithms of numeric (in exact arithmetic) deduction of
analytical expressions, proposed and described by~\cite{SV93}, are
developed and implemented in a computer algebra code. This code is
built as a superstructure for the computer algebra package
by~\cite{SS93a} for normalization of Hamiltonian systems of
ordinary differential equations, in order that high complexity
problems of normalization could be solved. As an example, a
resonant normal form of a Hamiltonian describing the hyperboloidal
precession of a dynamically symmetric satellite is derived by
means of the numeric deduction technique. The technique provides a
considerable economy, about 30 times in this particular
application, of computer's memory consumption. It is naturally
parallelizable. Thus the economy of memory consumption is
convertible into a gain in the computation speed.
\end{abstract}

\section{Introduction}

Complexity of symbolic computations usually depends on integer
parameters. E.g. in case of expanding a function in a series,
these complexity-governing parameters are the
expansion order and the number of variables;
in case of computing the determinant of a matrix with symbolic
elements, this parameter is the size of the matrix;
for solving a system of differential equations, this is the order
of the system; etc. Besides, the complexity of computations
depends on a number of free symbolic parameters in an
analytical expression under construction.

With an increase of $N$ (in what follows any complexity-governing
parameter) the complexity of computations usually also increases,
this growth however being different for an amount of intermediary
computations and for a volume of a final result. E.g. there may be
the following situation:
the amount of intermediary computations grows exponentially with $N$,
i.e. grows rapidly, but the final result has the complexity growth
linear in $N$, i.e. remains still simple.

Suppose that one resolves a computer algebra problem depending on
a parameter. Some variation, say growth, of the parameter leads
to growth of complexity of the computation. Let $V_{fin}$ be the
volume of the final result, $V_{int}$ be that of all intermediary
expressions. If the ratio $V_{fin}/V_{int}$ tends to zero with
increasing the value of the parameter, I call such a problem
in what follows a `generic parametric problem', since dependences
of $V_{fin}$ and $V_{int}$ on $N$ are expected to be
generically different.

An example of a problem analogously generic in the sense above,
but in ordinary non-symbolic computations, is provided by
computation of a chaotic trajectory of a dynamical system. The
exponential divergence of chaotic trajectories (see
e.g.~\citealt{LL92}) implies that a linear increase of accuracy of
the output of coordinates at a given time of the trajectory's
evolution requires an exponential increase of accuracy of starting
values and that of computation as a whole.

The term `generic' does not straightforwardly mean that
generic parametric problems are most abundant in applications
of computer algebra. This deserves a separate study.
Generic parametric problems often emerge
when it is necessary to accomplish analytical simplification
of intermediary expressions with respect to parameters. It
takes space and time. One of such examples,
concerning normalization of systems of ordinary differential
equations, is considered in this paper.

Rapid growth of complexity with increasing the parameter $N$ leads
to a fast exhaustion of computer's memory and impossibility of
further analytical computation. Radical means for economy of
memory consumption are provided by the method of numeric deduction
of analytical expressions, suggested by \cite{SV93}. It consists
in restoration of an analytical expression on a set of its exact
numeric evaluations obtained on a set of some simple exact numeric
values of parameters which the derived expression depends upon.
One should stress that {\it exact} arithmetic is used, not the
approximate one usually implied by `numeric computation'.

The method of numeric (or, to put it rigorously, exact-numeric)
deduction is essentially based on an extension to rational
functions of the `evaluation-interpolation' technique for
computations with polynomials. This latter technique is standard
in computer algebra (see e.g.~\citealt{GCL92}, Chapter~5).

The method of numeric deduction (or, the formula-guessing
technique) may constitute the only means
for deriving analytical expressions, when the parametric problem
is generic in the sense above and the value of $N$ is high.

The motivation of the following study is to investigate
the benefits of the formula-guessing technique in a real analytical
computation. The plan of the paper is as follows. First, theoretical
basics for the numeric deduction of analytical
expressions are recalled. Then the computer algebra package `Norma'
for normalization of Hamiltonian systems of ordinary differential
equations and the problem of normalization itself are
described briefly. Then a computer algebra code of numeric deduction,
written in the REDUCE language, is described.
Then, normalization of a resonant Hamiltonian for the
hyperboloidal precession of a symmetric satellite, accomplished
by means of this code, is considered. Finally, major results of the
experience in application of the code are analysed;
these results consist in economy of computer's memory and
opportunities for parallelization.

Indeed, the term `parallelization' provides a better
grasp on the essence of the method of numeric deduction.
This technique can thus be called the method of
`numeric parallelizing',
or, more rigorously, `exact-numeric parallelizing',
since computations in exact arithmetic are implied.

\section{
Theoretical basics for numeric deduction of analytical
expressions}
\label{basics}

Numeric deduction of an analytical expression consists in its
restoration upon a set of exact numeric values of parameters which
the expression depends upon. In this section the basics of this
method are recalled following~\cite{SV93}. Exclusively exact
arithmetic is implied in what follows.

Given the numeric data set computed for various values of
paramaters which the unknown analytical expression depends upon,
one may try to recover this expression. There are two main
ingredients in the procedure of numeric deduction: (1)~recovery of
structure of the derived expression from its numeric evaluations,
and (2)~Pad\'e interpolation of numeric values which have one and
the same location in this structure. A numeric data set subject
for restoration of an analytical expression may be e.g. of the
form $ 1+3*\mathrm{SIN}(1/3)$, $3/2+7/9*\mathrm{SIN}(5)$, etc.

The first major problem is that of distortion of structure of
restored expressions. Namely, there exist degenerate cases, when
the structure is distorted, because prefix forms, representing
transcendental and algebraic functions, disappear due to
simplification rules: e.g. $\mathrm{LOG}(1)=0$,
$\mathrm{SQRT}(4/9)=2/3$. What is more, the most hasardous for the
accomplishment of the procedure of restoration is the distortion
of the kind $\mathrm{SQRT}(5/9)=1/3*\mathrm{SQRT}(5)$, when
numbers `drift' from under the prefix. Probabilities of distortion
for prefix forms representing transcendental and algebraic
functions constitute an hierarchy. For some important functions,
the probabilities can be found in \cite{SV93}. E.g. the
probability of distortion for $\mathrm{SQRT}$ (square root) is
equal to $0.39$ when its argument is integer, and to $0.53$ when
it is rational; that for $\mathrm{CBRT}$ (cubic root) is equal to
$0.17$ and $0.23$ accordingly. These are average probabilities
calculated for argument values taken at random.

The second major problem is that of verification of a restored
expression. It can be verified

\noindent
(1) by means of an independent check
(e.g. a solution of an equation can be verified
by its direct substitution in the equation);

\noindent
(2) by proving analytically that the powers of restored
rational functions in the procedure of Pad\'e interpolation
have some upper bounds, and computing the sufficient number
of evaluations;

\noindent
(3) by checking the derived expression on an extra set of
numeric evaluations, and relying on the assumption that the
probability of an accident coincidence is zero.

In the example of normalization of a Hamiltonian system of
ordinary differential equations, considered in this paper,
the third way is chosen.
Besides, the resulting expression obtained by means of numeric
deduction, is independently derived by means of a direct symbolic
computation requiring much greater memory expenditures.

\section{Computer-algebraic normalization of Hamiltonian systems
of ordinary differential equations}

Reduction of a Hamiltonian system of ordinary differential equations
to a normal form is often used to derive an analytical solution of
the system or to analyse its stability. In particular, the method of
normal forms allows one to find approximate general solutions
in the neighbourhood of points of equilibria or periodic motions
and to analyse stability of motion in their neighbourhood.

The specialized application package `Norma'~\citep{SS93a} is
intended for an analytical accomplishment of procedures necessary
for normalization of autonomous Hamiltonian systems. The codes of
the package are written in the language of the REDUCE~3.2 computer
algebra system~\citep{H85,YY89}. The package allows one to
accomplish linear and non-linear normalization of the systems.
Besides, it is important that it utilizes a special
memory-consuming algorithm to derive expansion of a Hamiltonian in
power series with respect to canonical variables in the
neigbourhood of a fixed point.

Let $q_j$, $p_j$ be the coordinate and momentum variables; $j =
1$, $\ldots$, $N$, where $N$ is the number of degrees of freedom.
When all the eigenvalues of the matrix of a Hamiltonian system
linearized in the neighbourhood of a point of equilibrium are
strictly imaginary, and resonances up to the second order
inclusive are absent, i.e. there are no zero or equal frequencies,
the quadratic part of the Hamiltonian, according to~\cite{A74}, is
reducible to the normal form

\begin{equation}
K^{(2)}  = {1\over 2} \sum^{N}_{j=1} \lambda_{j}(q^{2}_{j} + p^{2}_{j}),
\label{h2}
\end{equation}

\noindent where $\lambda_{j} = \delta_{j}\omega_{j}$,
$\delta_{j} = \pm 1$.
The quantities $\omega_{j}=\vert \lambda_{j} \vert$ are
the frequencies of the linearized system.

In the non-resonant case, in the `polar' canonical variables
$r_j$, $\varphi_{j}$, defined by the formulas

\begin{equation}
q_{j} = \sqrt{2r_j} \, \sin \varphi_{j}, \
p_{j} = \sqrt{2r_j} \, \cos \varphi_{j},
\end{equation}

\noindent the Birkhoff normal form of order $M \ge 4$, according
to~\cite{A74}, is

\begin{equation}
K^{(M)} =
 \sum_{j=1}^N {\lambda_{j}r_{j}} + \sum_{n=2}^{[M/2]}
 \sum_{\ell_1+\ldots+\ell_N=n} c_{\ell_1, \ldots, \ell_N}
 r_1^{\ell_1} \ldots r_N^{\ell_N},
\label{BNF}
\end{equation}

\noindent
where $[M/2]$ is the round part of $M/2$. The form $K^{(M)}$
does not depend on angle variables. Note that forms $K^{(M)}$
and $K^{(M-1)}$, with odd $M \ge 3$, coincide. A Hamiltonian
normalized up to the order $M$ is given by the formula
$K = K^{(M)} + h^{(\ge M+1)}$, where $h^{(\ge M+1)}$ represents
terms of degree $M+1$ and higher with respect to the variables
$q_{j}$ and $p_{j}$ (or, equivalently, terms of degree
higher than $[M/2]$ with respect to the variables $r_{j}$);
these terms may depend on angle variables.

By definition, the resonance takes place if a set of integer
numbers $k_{j}$ exists such that

\begin{equation}
\sum  ^{N}_{j=1} k_{j}\omega_{j} = 0, \ \
{\bf k} = \sum^{N}_{j=1}\vert k_{j}\vert  \neq  0,
\end{equation}

\noindent
where $\omega_{j}$ are the frequencies, $k_{1} \ge  0$.
The quantity ${\bf k}$ is the order of the resonance.

On condition that there is no resonance of the kind
$k_{1}\omega_{1} + k_{2}\omega_{2} = 0$
(where ${\bf k} = k_{1} + \vert k_{2}\vert \le 6)$,
the normalized Hamiltonian of a system with two degrees
of freedom is

\begin{eqnarray}
&& \!\!\!\!\!\!\!\!\!\!\!\!\!
K = \lambda_1 r_1 + \lambda_2 r_2 + \\ \nonumber
&& \!\!\!\!\!\!\!\!\!\!\!\!\!
{} + c_{20}r^{2}_{1} + c_{11}r_{1}r_{2} + c_{02}r^{2}_{2} +
\\ \nonumber
&& \!\!\!\!\!\!\!\!\!\!\!\!\!
{} + c_{30}r^{3}_{1} + c_{21}r^{2}_{1}r_{2} +
c_{12}r_{1}r^{2}_{2} + c_{03}r^{3}_{2} +
\\ \nonumber
&& \!\!\!\!\!\!\!\!\!\!\!\!\!
{} + \mbox{(terms of higher order) },
\end{eqnarray}

\noindent
where $c_{20}$, $c_{11}$, $c_{02}$, $c_{30}$, $c_{21}$,
$c_{12}$, $c_{03}$ depend on parameters of the system.

In case of a resonance, the normalized Hamiltonian contains
additional terms which depend also on angle variables and
which cannot be eliminated. For the resonance
$k_{1}\omega_{1} + k_{2}\omega_{2} = 0$,
$k_{1} + \vert k_{2}\vert \ge 3$,
the normalized Hamiltonian of a system with two degrees of
freedom is

\begin{eqnarray}
&& \!\!\!\!\!\!\!\!\!\!\!\!\!
K = \lambda_1 r_1 + \lambda_2 r_2 +
c_{20}r^{2}_{1} + c_{11}r_{1}r_{2} + c_{02}r^{2}_{2} +
\label{rh}
\\ \nonumber
&& \!\!\!\!\!\!\!\!\!\!\!\!\!
{} + \mbox{(non-resonant terms of higher order) } +
\\ \nonumber
&& \!\!\!\!\!\!\!\!\!\!\!\!\!
{} + A_{k_1 k_2} {r}^{k_1 /2}_1 {r}^{\vert k_2 \vert /2}_2
sc(\delta_{1}k_{1}\varphi_{1} + \delta_{2}k_{2}\varphi_{2}) +
\\ \nonumber
&& \!\!\!\!\!\!\!\!\!\!\!\!\!
{} + \mbox{(resonant terms of higher order) },
\end{eqnarray}

\noindent
where $sc$ denotes sine or cosine,
$\delta_j = \pm 1$ are defined in Eq.~(\ref{h2}).
The quantities
$A_{k_{1}k_{2}}$, as well as $c_{\ell_1, \ldots, \ell_N}$
in Eq.~(\ref{BNF}), are invariants of the Hamiltonian
with respect to canonical normalizing transformations.

In the `Norma' package, the non-linear normalization is performed
by the method based on Lie transformations~\citep{H66,D69,M70}.
The number of degrees of freedom and the order of normalization
are arbitrary. The coefficients of the initial Hamiltonian may
have symbolic or exact numeric representation. The code of
non-linear normalization in the `Norma' package computes the
normalized Hamiltonian and the generating function of the
normalizing transformation.

\section{The code for numeric deduction of analytical expressions}

To show how the method of numeric deduction works, I apply the
`Norma' specialized package to studies of small--amplitude
periodic motions in the neighbourhood of regular precessions of a
dynamically symmetric satellite in a circular orbit around a point
gravitating mass. The precession is called hyperboloidal, when a
satellite's axis of symmetry describes a hyperboloidal surface in
space~\citep{B75}.

In what follows, an analytical expression for the normal form in
case of a particular resonance is derived. The final formula is
important for studies of the orbital stability of motion in the
neighbourhood of the hyperboloidal precession~\citep{SS95}. Note
that formulas for normal forms for the hyperboloidal and
cylindrical precessions for various resonant and non-resonant
cases can be found in \cite{SS93b,SS95}; expressions given there
were obtained by means of a direct symbolic computation. The
subject of consideration in what follows is not the final formula
itself, but the way of its deduction.

The direct symbolic computation~\citep{SS95} shows that the
analytical complexity of resonant normal forms in the problem of
hyperboloidal precession grows linearly with the order of
normalization, while the volume of intermediary calculations, due
to necessity of analytical simplification of intermediary
analytical expressions, almost doubles with every order of
normalization, i.e. grows exponentially. It means that the
computer algebra problem of normalization of the Hamiltonian for
the hyperboloidal precession is generic as defined in the
Introduction.

Algorithms of numeric deduction are implemented here
in a computer algebra code as a superstructure for the `Norma'
computer algebra package.
A specific code implementation for an individual problem
seems to be the most promising approach for applications
of the formula-guessing technique, since the variety
of possible applications is too great to attempt to build
a universal system.

According to Section~\ref{basics},
a code implementing the formula-guessing technique should
include two main parts: a part for the structure analysis
of numeric data, and that for numeric restoration of
`remnants' of analytical expressions.
The structure of restored expressions in the problem
under study is relatively simple. The code for its
analysis was written especially for this problem.
It does not have general significance and
is not described here.

The part realizing the Pad\'e interpolation is of a general
applicability. It implements restoration of rational functions
which produce elementary numeric remnants. Prefix forms
representing transcendental and algebraic functions, according to
Section~\ref{basics}, are generally much less destructible. The
language of the REDUCE computer algebra system~\citep{H85} was
used in writing this code of restoration of rational functions.
The set of numeric evaluations of the final expression, obtained
by means of consecutive application of the `Norma' package for a
set of numeric values of the parameters of the problem, serves as
an input for the code.

First the code checks if the number of an expression evaluations
is sufficient for restoration of a rational function with
prescribed lengths of polynomials in the numerator and denominator.
The lengths are specified by setting the minimum and maximum values
of the degrees of terms in the interpolating polynomials.
They are designated in the code by $k$ and $l$ for the numerator,
and by $m$ and $n$ for the denominator.
Before computation, certain assumptions can be made
on the possible ranges of $k$, $l$, $m$, $n$.
E.g., in the example of the code given below, it is assumed
that the denominator consists of a single monomial of a prescribed
degree (guessed by induction from the appearance of coefficients
of lesser order). Taking wider ranges would result in a somewhat
greater computation time.
If one makes a wrong assumption on these ranges, the procedure
either complains on the insufficiency of data, when the number
of data points is insufficient; or fails to
verify the final expression on the additional data set,
when the assumed ranges do not cover real ones.
The data are insufficient, if $n_{sum} = l-k+n-m+1$ is greater
than the number of data points.

Generally, when no assumptions are made beforehand on the ranges
of $k$, $l$, $m$, $n$, the algorithm is as follows. The values
of $k$, $m$ are set to zero; the values of $l$, $n$ are
step by step increased from zero, and for each set of
$k$, $l$, $m$, $n$, the rational function is restored.
Thus the values of $l$, $n$ are increased until the rational
function does not change anymore, i.e. its form stabilizes,
and it fits all data points. If at some step the data are
insufficient, more numeric data points are computed and
added to the data set. Note that the time and memory expenditures
at the stage of restoration are negligible in comparison with
main expenses, which are associated with the construction of
the numeric data set.

In the example which follows, after checking the sufficiency of
data, the input data are squared, since the resulting expression,
judging from its numeric remnants presented below, contains
square roots. Then the procedure $rfn$ of the Pad\'e
interpolation is called. It produces the resulting expression $f$,
and the latter is verified for the remaining data points.

The procedure $rfn$, implementing the Pad\'e interpolation,
restores a rational function $f$ from its numeric remnants.
Undetermined coefficients and some linear algebra are used.

Normal forms of the Hamiltonian for the hyperboloidal precession
are found by means of application of these procedures to
the data obtained beforehand by means of consecutive application
of the `Norma' package to a set of numeric values of a parameter
of the problem. In the following example, these data are obtained
for the case of the resonance $\omega_1 = 5 \omega_2$ between
the frequencies of the system. An extract from the file with
the data is given below. The designations are:
$x(i)$ is a numeric value of the frequency $\omega_2$
of the system, $y(i)$ is the evaluation of the resonant normal
form at this point, $i$ enumerates the points, $npoints$ is their
total number.

\smallskip
\smallskip
\smallskip

$$
\begin{array}{rcl}
&& \!\!\!\!\!\!\!\!\!\!
   \mbox{npoints}:=23;              \\
&& \!\!\!\!\!\!\!\!\!\!
   \mbox{x}(1):=19/104*\mbox{sqrt}(19)**( - 1)*
   \mbox{sqrt}(26);        \\
&& \!\!\!\!\!\!\!\!\!\!
   \mbox{y}(1):=901287283/454115447307648*      \\
&& \!\!\!\!\!\!\!\!\!\!
   \mbox{sqrt}(5)*\mbox{sqrt}(19)**( - 1)*       \\
&& \!\!\!\!\!\!\!\!\!\!
   \mbox{sqrt}(26)**( - 1)*\mbox{sqrt}(6726)*    \\
&& \!\!\!\!\!\!\!\!\!\!
   \mbox{sqrt}(45258)*\mbox{sqrt}(\mbox{R}(1))*
   \mbox{sqrt}(\mbox{R}(2))*            \\
&& \!\!\!\!\!\!\!\!\!\!
   \mbox{R}(2)**2*\cos(5*\mbox{FI}(2) -
   \mbox{FI}(1));                 \\
&& \!\!\!\!\!\!\!\!\!\!
   \cdot \cdot \cdot \cdot \cdot \cdot
   \cdot \cdot \cdot \cdot{}         \\
&& \!\!\!\!\!\!\!\!\!\!
   \mbox{x}(23):=83/104*\mbox{sqrt}(13)*
   \mbox{sqrt}(83)**( - 1);       \\
&& \!\!\!\!\!\!\!\!\!\!
   \mbox{y}(23):= - 10727690489953879/               \\
&& \!\!\!\!\!\!\!\!\!\!
   41357946769086552192*\mbox{sqrt}(5)*               \\
&& \!\!\!\!\!\!\!\!\!\!
   \mbox{sqrt}(13)**( - 1)*\mbox{sqrt}(83)**( - 1)*       \\
&& \!\!\!\!\!\!\!\!\!\!
   \mbox{sqrt}(373002)*\mbox{sqrt}(619014)*\mbox{sqrt}
   (\mbox{R}(1))*         \\
&& \!\!\!\!\!\!\!\!\!\!
   \mbox{sqrt}(\mbox{R}(2))*\mbox{R}(2)**2*
   \cos(5*\mbox{FI}(2) - \mbox{FI}(1));      \\
&& \!\!\!\!\!\!\!\!\!\!
   \mbox{end};                             \\
\end{array}
$$

\smallskip
\smallskip
\smallskip

\noindent
where R and FI correspond to $r$ and $\varphi$ in Eq.~(\ref{rh}).
The quantities $x(i)$ and the numeric coefficients of $y(i)$
in this data set are squared and forwarded to the
restoration procedure $rfn$; i.e. the functional part of $y(i)$,
which depends on R(1), R(2), FI(1), FI(2),
and is one and the same for all points, is set to unity,
because one is interested only in numeric coefficients of
$y(i)$.

By means of application of the procedure $rfn$ of the Pad\'e
interpolation, one gets the analytical expression $f$ (which
is the square of the desired expression):

\smallskip
\smallskip
\smallskip

$$
\begin{array}{rcl}
f:=( \!\!\!\!\!\!\!\!\!\!\!\!
&& {} - 117205809409155600*s**12   \\
&& {} + 324914084622543024*s**11   \\
&& {} - 335312660614677372*s**10   \\
&& {} + 161733011003713812*s**9    \\
&& {} - 39226577139649249*s**8     \\
&& {} + 5576587050768892*s**7      \\
&& {} - 508513621896676*s**6       \\
&& {} + 31144123897436*s**5        \\
&& {} - 1302165401582*s**4         \\
&& {} + 36818043284*s**3           \\
&& {} - 675424552*s**2             \\
&& {} + 7273552*s                  \\
&& {} - 34969)/(26759446470328320*s**13);  \\
\end{array}
$$

\smallskip
\smallskip
\smallskip

\noindent where $s = \omega_2^2$. The frequency $\omega_2$ is
expressed through the initial parameters of the problem
(see~\citealt{SS95}). After factorization and taking square root
of $f$, the final expression for the resonant coefficient $A_{1,
-5}$ is obtained:

$$
\!\!\!\!\!\!\!\!\!\!\!\!\!\!\!\!\!\!\!\!\!\!\!\!\!\!\!\!\!\!
\!\!\!\!\!\!\!\!\!\!\!\!\!\!\!\!\!\!\!\!\!\!\!\!\!\!\!\!\!\!
\!\!\!\!\!\!\!\!\!\!\!\!\!\!\!\!\!\!\!\!\!\!\!\!\!\!\!\!\!\!
\!\!\!\!\!\!\!\!\!\!\!\!\!\!\!\!\!\!\!\!
 A_{1,-5} = {(1 - s)^{1/2}(25 s - 1)^{1/2}(21 s - 1)
            \over 73156608*5^{1/2} s^{13/2}}*{}
$$
\begin{equation}
\qquad \qquad \qquad \qquad
 {}*(3260508 s^{4} - 2668610 s^{3} + 312005 s^{2} - 13090 s + 187).
\end{equation}

The described above application of the formula-guessing technique
allows one to deduce the needed expression for the resonant
coefficient, when not more than 100 Kb is provided
for storage during computations.
This expression can also be obtained by an ordinary direct symbolic
computation, when much greater memory is allowed to consume.
If one uses a direct symbolic computation,
intermediary analytical expressions in the procedure of analytical
non-linear normalization occupy megabytes of memory, but the
final expression is compact enough to be presented, as it has been
just shown, in typographical form.

This direct symbolic computation turned out to require
more than 3 Mb of memory,
i.e. more than 30 times greater than in case of the
numeric deduction computation. The reason for this economy is clear:
it needs much less space to store numerals, than to store complicated
analytical expressions. Besides, it takes less time and memory
to operate with numerals than with analytical formulas.
The formula-guessing technique uses very small amount of
memory for computations
at each point (at each numeric starting value) in the data set.
As the computation is made in a sequence, from point to point,
the over-all economy of memory is achieved.

This gain in memory's economy can be converted
into a gain of the computation speed, if the computation on the
data set is performed not in a sequence, but in parallel.
Indeed, the final procedure of restoration of an expression
is not time-consuming; it is the construction of the data set
for this restoration which constitutes a major part of time
expenditures.
If initial data sets, i.e. sets of numeric values of parameters,
are constructed to be of homogeneous complexity,
the elementary processes of numeric evaluations of the desired
expression at each point
would be approximately of one and the same duration.
This means that the advantage in memory consumption is convertible
into an advantage in the computation speed,
if numeric evaluations of the resulting expression on the set of
parameters' values are performed in parallel.

\section{Conclusions}

The major benefit of using the formula-guessing
technique, in comparison with the direct symbolic computation,
consists in a considerable decrease of computer's
memory consumption. As it was shown above in the example
of application of this technique to the problem of normalization
of a Hamiltonian system of ordinary differential equations,
the relative decrease in the memory consumption can be more
than 30 times. This advantage in the memory's economy is convertible
into a gain of computation speed.
This follows from the fact that the elementary processes of numeric
deduction are approximately of one and the same duration, if
the set of initial data is constructed to be of homogeneous complexity.
Thus the algorithm of numeric deduction of analytical
expressions is naturally parallelizable.

\bigskip
It is a pleasure to thank Andrej Sokolsky and Nikolay Vasiliev
for useful discussions.

{}

\end{document}